\documentclass[conference]{IEEEtran}
\IEEEoverridecommandlockouts
\usepackage{cite}
\usepackage{amsmath,amssymb,amsfonts}
\usepackage{algorithmic}
\usepackage{graphicx}
\usepackage{booktabs}
\usepackage{multirow}
\usepackage{textcomp}
\usepackage{siunitx}
\usepackage{xcolor}
\usepackage{soul, color}
\usepackage{algorithm}
\usepackage{booktabs}
\usepackage{fancyhdr}
\def\BibTeX{{\rm B\kern-.05em{\sc i\kern-.025em b}\kern-.08em
    T\kern-.1667em\lower.7ex\hbox{E}\kern-.125emX}}
\begin{document} 

\title{Machine Learning--Based Protection and Fault Identification of 100\% Inverter-Based Microgrids\\

\thanks{
© 20XX IEEE.  Personal use of this material is permitted.  Permission from IEEE must be obtained for all other uses, in any current or future media, including reprinting/republishing this material for advertising or promotional purposes, creating new collective works, for resale or redistribution to servers or lists, or reuse of any copyrighted component of this work in other works.

This work of Virginia Tech is supported in part by the National Science Foundation (NSF) under award ECCS-1953213, in part by the State of Virginia’s Commonwealth Cyber Initiative (www.cyberinitiative.org), in part by the U.S. Department of Energy's Office of Energy Efficiency and Renewable Energy (EERE) under the Solar Energy Technologies Office Award Number 38637 (UNIFI Consortium led by NREL), and in part by Manitoba Hydro International. The views expressed herein do not necessarily represent the views of the U.S. Department of Energy or the United States Government. The work of UAA is supported in part by the U.S. National Science Foundation (NSF) under awards OIA-2229772, RISE-2220624, RISE-2022705, and RISE-2318385.
}
}

\author{
\IEEEauthorblockN{Milad Beikbabaei,$^1$ \emph{Graduate Student Member, IEEE}, Michael Lindemann,$^2$ \emph{Graduate Student Member, IEEE},\\ Mohammad Heidari Kapourchali,$^2$ \emph{Member, IEEE}, and Ali Mehrizi-Sani,$^1$ \emph{Senior Member, IEEE}}
\IEEEauthorblockA {$^1$ The Bradley Department of Electrical and Computer Engineering, Virginia Tech, Blacksburg, VA 24061 \\
\IEEEauthorblockA{$^2$ Department of Electrical Engineering, University of Alaska, Anchorage, AK 99508}
e-mails: miladb@vt.edu, mlindemann@alaska.edu, mhkapourchali@alaska.edu, mehrizi@vt.edu }
}

\maketitle

\begin{abstract}
100\% inverter-based renewable units are becoming more prevalent, introducing new challenges in the protection of microgrids that incorporate these resources. This is particularly due to low fault currents and bidirectional flows. Previous work has studied the protection of microgrids with high penetration of inverter-interfaced distributed generators; however, very few have studied the protection of a 100\% inverter-based microgrid. This work proposes machine learning (ML)--based protection solutions using local electrical measurements that consider implementation challenges and effectively combine short-circuit fault detection and type identification. A decision tree method is used to analyze a wide range of fault scenarios. PSCAD/EMTDC simulation environment is used to create a dataset for training and testing the proposed method. The effectiveness of the proposed methods is examined under seven distinct fault types, each featuring varying fault resistance, in a 100\% inverter-based microgrid consisting of four inverters. 
%(a) General introduction and statement of the problem;
%(b) What has been done and their shortcomings;
%(c) What is proposed in this paper (very clearly, probably as a bullet list) and its merits as compared with the previous paragraph;
%(d) Methodology adopted for the proposed method;
%(e) Major conclusions and outcomes; and
%(f) Structure of the rest of the paper.

\end{abstract}


\section{Introduction}

With the increasing prevalence of renewable energy resources in the form of inverter-based distributed generation units, traditional protection strategies for microgrids connecting multiple such resources may become insufficient. The protection of 100\% inverter-based microgrids presents significant challenges~\cite{sharma2023new}, primarily due to the reduced inertia of inverter-based units~\cite{milad_FDI_LSTM}. %mishra2015combinedz 
A microgrid with low inertia sources could experience stability problems if line faults are not quickly cleared. In inverter-interfaced distributed generators, the current contribution is limited during short circuits, resulting in much lower fault currents. Limited fault current, combined with the bidirectional power flow and intermittent generation, presents a challenge for fault protection when employing traditional high fault current methods.

%\hl{Previous work has proposed protection schemes in a grid with a high-penetration of inverter-based resources (IBR).}
%The authors in~\cite{enhancing_current_SC} suggest co-locating inverter-based resources (IBR) with synchronous condensers to enhance overcurrent protection comparable to synchronous generator-based resources, thereby improving the accuracy of the protection system.
%
%A protection scheme is developed in~\cite{Mehrizi_Leastsquare} by comparing the measured current with the estimated current; however, it cannot detect fault type.
%
Reference~\cite{adaptive_coordination_inverter_based} studies the challenges faced when converting an existing distribution feeder to an inverter-based microgrid and suggests the use of adaptive settings for relays to tackle the protection challenges. Protection challenges are more prominent in isolated microgrids, which have lower fault levels compared to grid-connected microgrids with infeed power from the main grid~\cite{liu2022transient}. A protection scheme for isolated microgrids with high penetration of inverter-interfaced distributed generators should be fast, adaptable, and accurate in order to uphold microgrid stability and safeguard critical loads. However, the research on microgrid protection has not yet led to a commercially available microgrid relay~\cite{kandasamy2023intelligent}. This is because many solutions rely on communication methods or complex learning-based relay systems~\cite{gadde2021topology}. 
Using communication makes the power system susceptible to delays and cyberattacks, reducing the grid's resiliency~\cite{milad_API_PSCAD}. 
Communication delays can affect the dynamic behaviors of communication-based IBR control during faults, which makes protection more challenging~\cite{communication_delay_milad}.
Many existing microgrid protection schemes lack adaptability to diverse topologies and source types, are not cost-effective, or rely on communication, and have not been tested under scenarios with extremely high IBR penetration, which is likely in the near future~\cite{gadde2021topology}.

Machine learning (ML)–based protection methods have shown great potential in accurately detecting and identifying faults. However, they have been mostly employed in transmission and distribution systems protection, and very few studies have explored the applications of ML in detecting and identifying faults in microgrids with high renewable resource penetration~\cite{kandasamy2023intelligent}. 
Authors in~\cite{kandasamy2023intelligent} propose an intelligent fault diagnosis method based on deep learning, utilizing wavelet transformation and sequence components. Deep learning models developed in Pytorch are employed to train and validate fault detection, classification, and location identification. 
Authors in~\cite{aiswarya2023novel} present a general-purpose support vector machine (SVM)-based adaptive scheme to identify normal and fault conditions in AC microgrids and detect fault types. Reference~\cite{uzair2022machine} develops an ML-based protection method for AC microgrids that detects and classifies faults. %In~\cite{uzair2022machine}, the random forest technique outperformed all other ML classifiers. 
The proposed ML methods use complex deep learning algorithms or require high sampling rates. 

This paper presents decision tree-based protection solutions that combine fault detection and fault type classification in a fully inverter-based microgrid, using local measurements without any communication.
The effectiveness of the protection solution is studied on a microgrid equipped with four inverters, where half operate in grid-following mode, and the remaining two function in grid-forming mode. 
%In this study, different ML-based protection method\textcolor{red}{s} are implemented and compared under different short-circuit fault scenarios. 
The proposed method only uses the root mean square (RMS) value of the three-phase current, three-phase voltage, active and reactive power with 1~ms intervals, which reduces the computational process and memory use. The low computational burden facilitates practical implementation on a microcontroller.
The salient features of the proposed method are:
\begin{itemize}
\item The proposed method, designed for simplicity, integrates fault detection and identification for efficient implementation in digital relays.
\item The decision tree algorithm trains in less than 9~s makes it applicable for real-world grid applications.
\item The proposed approach can detect faults in less than 5~ms for both low- and high-impedance faults.
\end{itemize}

%Section~II presents an overview of the test system. Section~III discusses the methodology, Section~IV details the simulation results, and Section~VI concludes the paper.

\section{Test System }

\subsection{Basics of Inverter Control}

This subsection discusses the basics of inverter control, where it can be controlled either in the grid-following or the grid-forming mode~\cite{milad_FDI_LSTM}.

\subsubsection{Grid-Following Mode}

\begin{figure}[!t]
%\begin{figure}[htbp]
\centerline{\includegraphics[width= 0.9\columnwidth]{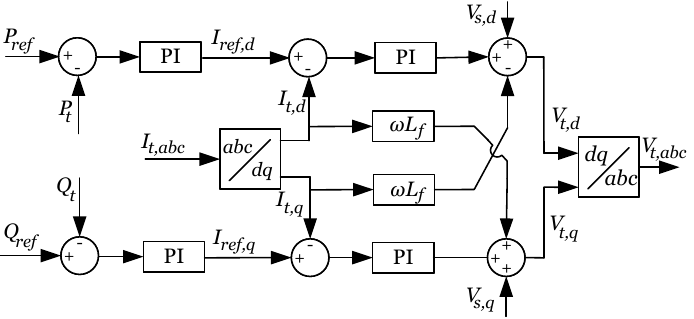}}
 %\vspace*{-0.18 cm}
\caption{Decoupled control scheme of a grid-following inverter for $d$- and $q$-axes.}
\label{decoupledCurrentControl}
\end{figure}

The inverter receives real power set points $P_\text{ref}$ and reactive power set points $Q_\text{ref}$, and the inverter adjusts its output power to be as close as possible to the received power set points.
%In the grid-following mode, the inverter injects a certain amount of real and reactive power to the grid. 
A phase-locked loop (PLL) is used for estimating the voltage phase angle, helping with converting the current and voltage form $abc$-frame to $dq$-frame and back.
Fig.\ref{decoupledCurrentControl} shows the grid-following conventional decoupled current control loop for an inverter connected to the grid through an RL filter, where the output real power is $P_\text{t}$, the output reactive power is $Q_\text{t}$, inverter current is $i_{t}$, and $V_{t}$ and $V_{s}$ are the terminal and the grid voltages, respectively~\cite{GM_Naser}.
% The inverter control block in the grid-following mode is shown in Fig.~\ref{decoupledCurrentControl}.

\subsubsection{Grid-Forming Mode}
 
\begin{figure}[!t]
%\begin{figure}[htbp]
\centerline{\includegraphics[width= 0.9\columnwidth]{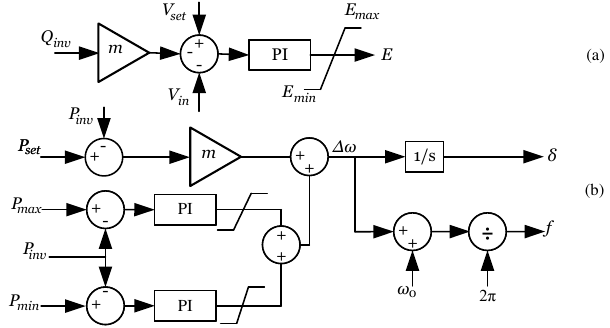}}
 %\vspace*{-0.18 cm}
\caption{Droop control scheme of a grid-forming inverter for: (a) $P-f$, (b) $Q-V$.}
\label{droop_control_GM}
\end{figure}

 The voltage and frequency of the grid are subject to disturbances and changes; however, the voltage and frequency magnitude need to be maintained within nominal ranges. As a result, $Q$$-$$V$ and $P$$-$$f$ droop controls are used in grid-forming inverters.
%In the grid-forming mode the inverter helps to maintain the voltage and frequency of the grid using $Q$$-$$V$ and $P$$-$$f$ droop controls.
 The $Q$$-$$V$ droop control updates the voltage set point of the inverter controller, resulting in adjusting the reactive power to maintain the voltage as shown in Fig.~\ref{droop_control_GM}(a).
 The $P$$-$$f$ droop control updates the voltage angle of the inverter, resulting in modifying the real power output to maintain the frequency as shown in Fig.~\ref{droop_control_GM}(b).
 The PI controllers are used in the grid-forming inverter droop control, as shown in Fig.~\ref{droop_control_GM}.

\subsection{100\% Inverter-Based Microgrid}

\begin{figure}[!t]
\centerline{\includegraphics[width=0.9\columnwidth]{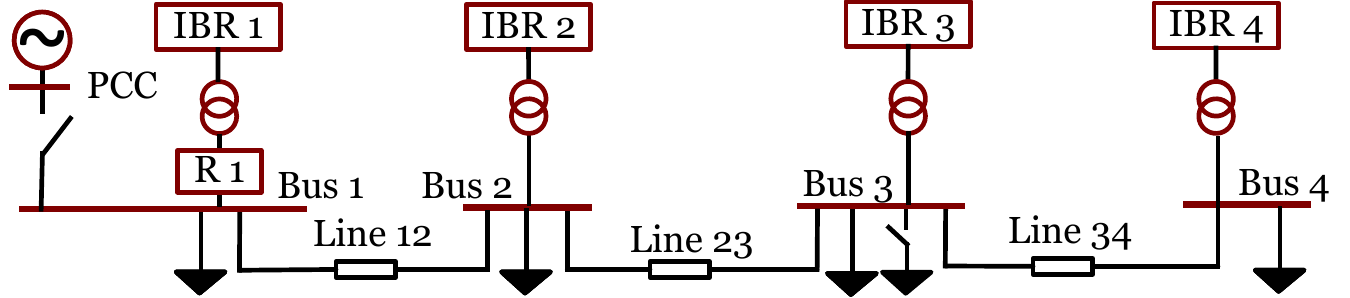}}
 %\vspace*{-0.18 cm}
\caption{ Single-line diagram of the microgrid.}
\label{GM_microgrid}
\end{figure}

A 4-bus microgrid is developed in PSCAD/EMTDC as shown in Fig.~\ref{GM_microgrid}, where the first and third inverters are grid-forming, and the rest are grid-following. The microgrid base power is 1.5~MVA, the low-voltage base voltage is 480~V, and the medium-voltage base voltage is 12.47~kV. Each inverter has a DC voltage source where its primary side voltage is fixed to 1.2~kV.
The maximum output power of each inverter is limited to 1.5~MVA. Each grid-following unit is connected to a transformer through an $RL$ filter, where its resistance is 1.5~m$\Omega$, and its inductance is $\SI{20}{\micro \text{H}}$.
$k_P=2$ and $k_I=0.0025$ are the gains for the grid following PI blocks.
Gains for the PI block in $P$$-$$f$ droop control of grid-forming inverters are $k_P=0.6$ and $k_I=0.003$, and they are $k_P=0.6$ and $k_I=0.002$ for the $Q$$-$$V$ droop control. 
Line~12 resistance is 1.4~$\Omega$, and line~12 inductance equals to 2~$\Omega$. 
Line~23 resistance is 2.2~$\Omega$, and line~23 inductance equals to 3.16~$\Omega$. 
Line~34 resistance is 0.6~$\Omega$, and line~34 inductance is 3.16~$\Omega$. 
A load is connected to bus~3 through a breaker, where its real power can be between 0.1~pu to 0.6~pu, and its reactive power can vary from 0.01~pu to 0.06~pu. The protection relay R1 is at bus~1.
%R1 is the relay located at bus~1.

\section{Methodology}

This work utilizes a decision tree--based method to detect and identify faults in the microgrid, as elaborated in the following subsections.

\subsection{Algorithm Selection}
Since the protection algorithm needs to detect a fault within a few milliseconds 
it needs to be run every few milliseconds. Due to this requirement, making algorithm less computational expensive is preferable. Previous work has utilized deep learning methods for protection; however, real-time implementation using off-the-shelf microcontrollers is still a challenge. Furthermore, microgrids are small grids with a few nodes, and classification algorithms can show effective results for protection against short circuit faults. The most basic type of algorithm is the support vector machine (SVM) model. However, SVM models scale between either quadratically or cubically with training data complexity. Due to this scaling penalty and this projects datasets complexity, SVM was not a viable option. 
% \hl{A support vector machine (SVM) model with a non-linear kernel scales either quadratically or cubically with the data complexity, and due to the large size of our data set, it shows insufficient results.}
Therefore, this work uses a decision tree model since decision tree models scale linearly with the data complexity. 
%Furthermore, the decision tree model shows high accuracy for both fault detection and fault type identification.

\subsection{Decision Tree Basics}

Decision trees work via a tree like structure in which each node represents a feature or attribute that has been determined as important. Branches correspond to a decision based on the previous feature and lead to the next node.
The decision on when to split in the tree is decided using the Gini impurity method. During training, the goal is to constantly minimize the Gini impurity~\cite{DT_ref}. The Gini impurity of the dataset is calculated using the equation:
%\[
%\text{Gini}(D) = 1 - \sum_{i=1}^{J} p_i^2,
%\]
\begin{align}
\begin{split}
\text{Gini}(D) = 1 - \sum_{i=1}^{J} p_i^2,
    \end{split}
    \label{eq4_GRU}
\end{align}
where $D$ represents the dataset for the node, $J$ is the number of classes, and $p_i$ is the proportion of samples that belong to class $i$ in the dataset $D$.

\subsection{ Dataset Preparation and Feature Selection}

A dataset is developed by simulating short-circuit faults using PSCAD/EMTDC. The PSCAD automation Python API is utilized to run multiple cases and create the dataset~\cite{milad_API_PSCAD}.
For each simulation run, a distinct fault type, fault location, fault resistance, and fault duration are considered. The seven fault types selected are $AG$, $BG$, $CG$, $ABG$, $ACG$, $BCG$, and $ABCG$.
The chosen fault resistance values are 100~$\Omega$, 10~$\Omega$, 1~$\Omega$, 0.1~$\Omega$, and 0.001~$\Omega$.
The faults are located at buses~1--4.
For the fault duration, 0.05~s, 0.1~s, and 0.2~s are selected. 
%For the fault duration, 0.05~s, 0.1~s, and 0.2~s are selected.
All faults are introduced at $t=0.05$~s, and each simulation is conducted for a duration of 1 second. As a result, a total of 420 cases are simulated and used for both training and testing the proposed protection algorithm.

\begin{table}[!t]
    \footnotesize\centering
    \caption{Fault Type Number Table}
    \label{tbl:type_fault}
    \begin{tabular}{llll}
        \toprule   
    \textbf{ Fault Number } & \textbf{ Description }& \textbf{ Fault Number } & \textbf{ Description }\\
        \midrule
        0 & No faults & 4 & $ABG$ faults\\
        1 & $AG$ faults & 5 & $ACG$ faults\\
        2 & $BG$ faults & 6 & $BCG$ faults\\
        3 & $CG$ faults & 7 & $ABCG$ faults  \\
        \bottomrule
    \end{tabular}
\end{table}

In every simulation, the three-phase current $I$, three-phase voltage $V$, real power $P$, and reactive power $Q$ data are recorded with a 1~ms time intervals.
Furthermore, a fault detection signal is added to the dataset, where it becomes 1 during the short-circuited faults and is zero when there is no fault. Additionally, a column is added for the fault type. A number is assigned to each fault type as shown in Table~\ref{tbl:type_fault}, where it can be between 0 and 7. Fault type 0 represents no faults.

\subsection{Training and Hyperparamters Selection}

\begin{table}[!t]
    \footnotesize\centering
    \caption{Table of the Accuracy of Different Input Features}
    \label{tbl:feature_selection}
    \begin{tabular}{llll}
    \toprule
     \multirow{2}{*}{\parbox{2cm}{\textbf{Inputs}}} &  \multirow{2}{*}{\parbox{2cm}{\textbf{Fault Prediction Accuracy}}} &  \multirow{2}{*}{\parbox{2.8cm}{\textbf{Fault Type Prediction\\ Accuracy}}} \\\\
    \midrule
          $I$ &                   95.58\% &                        95.50\% \\
        $I$,$V$ &                   95.92\% &                        95.82\% \\
      $I$,$P$,$Q$ &                   96.26\% &                        96.05\% \\
    $I$,$V$,$P$,$Q$ &                   96.33\% &                        96.04\% \\
    \bottomrule
    \end{tabular}
\end{table}

Table~\ref{tbl:feature_selection} shows how different selections of the input features affect the accuracy of the decision tree.
Four combinations of inputs are selected for the R1 relay, where the first one only uses $I$. The second one uses both $V$ and $I$. The third one uses $I$, $P$, and $Q$. The fourth one uses $I$, $V$, $P$, and $Q$. 
In order to achieve the highest accuracy the fourth feature set are used in this work, where it shows the best performance for fault detection compared to other combinations.
%
%
%Four combinations of inputs are selected for the R1 relay, where the first one only uses the three-phase current. The second one uses both three-phase voltage and current. The third one uses three-phase current, real power, and reactive power. The fourth one uses three-phase voltage and current and real and reactive power. 
%In order to achieve the highest accuracy the fourth feature set are used in this work, where it shows the best performance for fault detection compared to other combinations.

Scikit-learn library is used in this work, where the decision tree model maximum depth is set to 43 and it has a total of 4044 leaves.
Training and validation data consisted of 715,959 1~ms time samples. Input variables consisted of the three-phase current, the three-phase voltage, the real power output, and the reactive power output of the bus where the relay is located. The data is split utilizing \verb!train-test-split! function in scikit-learn. 80 percent of the data is used for training, and the remaining 20 percent is used for validation. A shuffle is applied with a random state setting of 20 being utilized for repeatability. No further preprocessing is required before training the decision tree model.
True output variables are split into their respective categories and fed into the model.

\section{Simulation results and Evaluation}

This section presents the simulation results for different types of faults at different buses.

\subsubsection{ Case~1: $AG$ Fault on Bus~1.}

\begin{figure}[!t]
\centerline{\includegraphics[width=0.9\columnwidth]{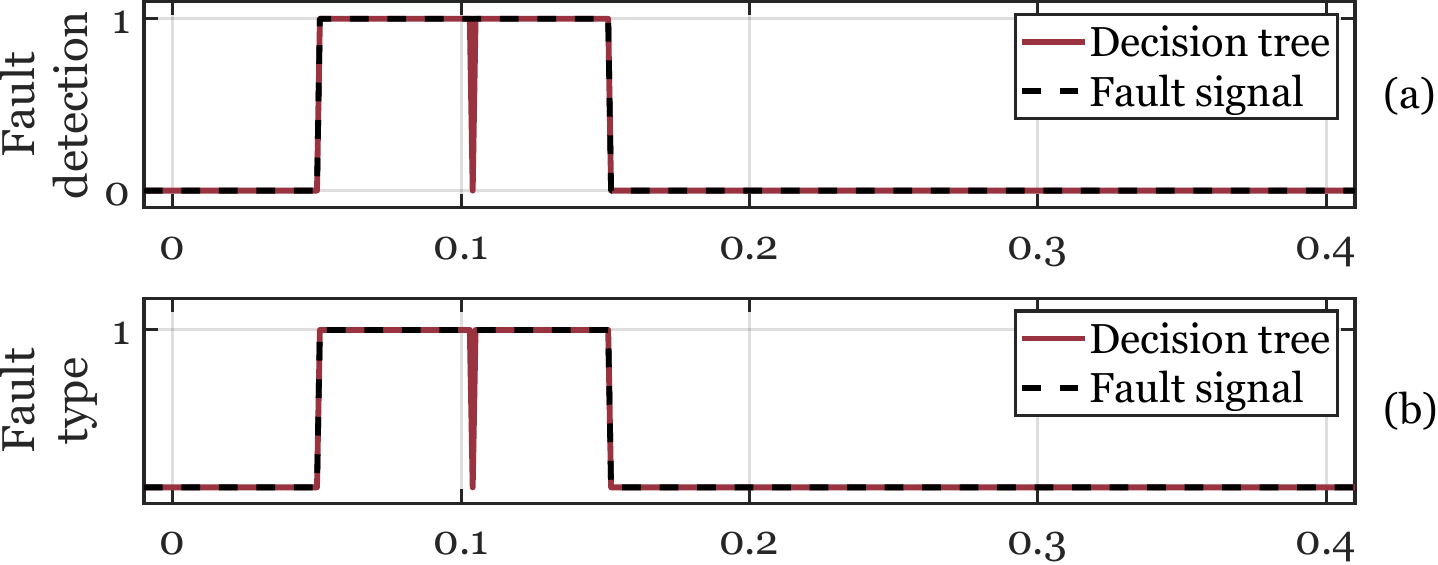}}
 %\vspace*{-0.18 cm}
\caption{ $AG$ fault located at bus~1: (a) fault detection and (b) fault type.}
\label{case1}
\end{figure}

%Figs.~\ref{case1}(a)--(c) show fault detection, fault type, and RMS current of relay for a phase $A$ to ground fault with a 0.01~$\Omega$ resistance located at bus~1, staring at $t=50$~ms and clearing at $t=150$~ms.
Figs.~\ref{case1} shows fault detection and fault type for a phase $A$ to ground fault with a 0.01~$\Omega$ resistance located at bus~1, staring at $t=50$~ms and clearing at $t=150$~ms.
The decision tree detects the fault occurrence at $t=50$~ms and detects fault clearance at $t=155$~ms, with 5~ms delays. The decision tree detects the fault type number 1, $AG$ fault.
The decision tree detects the fault type and fault occurrence correctly during the fault period except for one sample at $t=0.104$~s; however, it quickly corrects its prediction in the next time step. 
%Fig.~\ref{case1}(a) shows the RMS current of the relay at bus~1, where phase $A$ shows the highest increase compared to the other phases since the fault is $AG$.
%
%
%The decision tree estimates the fault impedance accurately.%

\subsubsection{ Case~2: $ACG$ Fault on Bus~2.}

\begin{figure}[!t]
\centerline{\includegraphics[width=0.9\columnwidth]{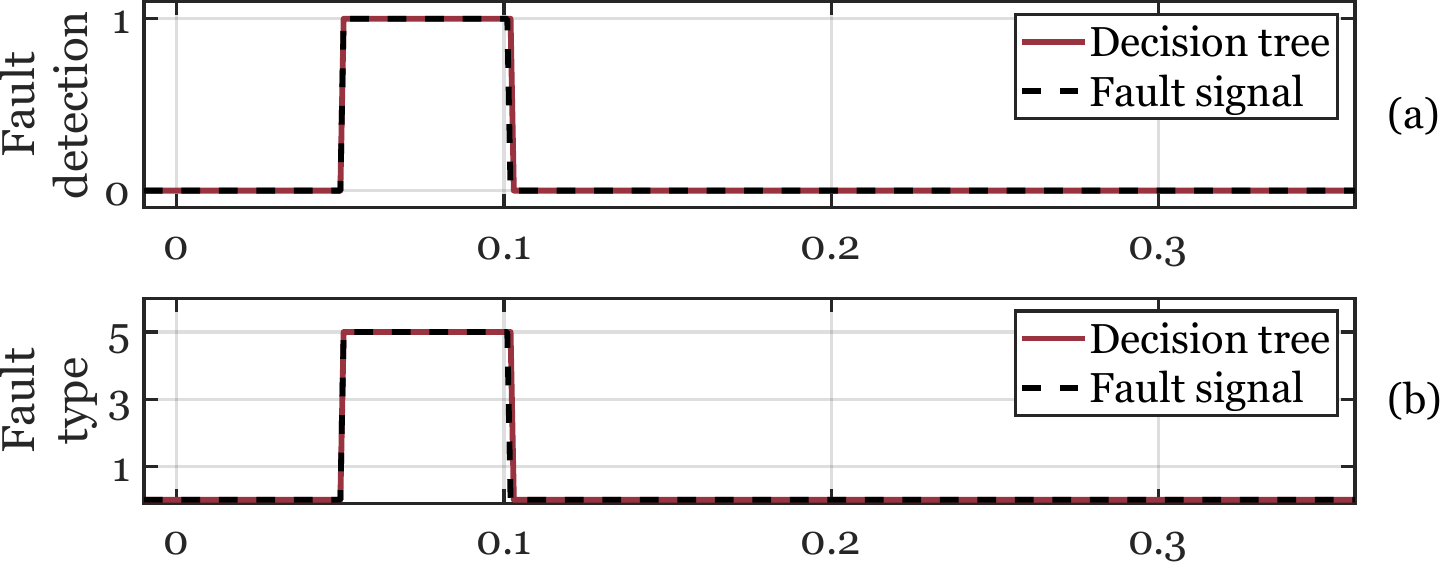}}
 %\vspace*{-0.18 cm}
\caption{ $ACG$ fault located at bus~2: (a) fault detection and (b) fault type.}
\label{case2}
\end{figure}

Figs.~\ref{case2} shows fault detection and fault type for phases $A$ and $C$ to ground fault with a 1~$\Omega$ resistance located at bus~2, starting at $t=50$~ms and clearing at $t=100$~ms.
%Figs.~\ref{case2}(a)--(c) show fault detection, fault type, and RMS current of relay for phases $A$ and $C$ to ground fault with a 1~$\Omega$ resistance located at bus~2, starting at $t=50$~ms and clearing at $t=100$~ms.
%
The decision tree detects the fault occurrence at $t=50$~ms and fault clearance at $t=100$~ms. 
The decision tree detects the fault type correctly.
The decision tree indicates that the fault type number is 5, which is equivalent to an $ACG$ fault  shown in Fig.~\ref{case2}(b).
%Fig.~\ref{case2}(b) shows the fault type number where the decision tree output is 5 during the fault, which is an $ACG$ fault using Table~\ref{tbl:type_fault}. 
%The decision tree detects the fault type correctly during the fault period except for one sample at $t=0.117$~s.
%Fig.~\ref{case2}(c) shows the relay current, where phases $A$ and $C$ currents increase during the fault due to the $ACG$ fault; however, all three-phase currents return to their pre-fault values after the fault is cleared.

\subsubsection{ Case~3: $ABCG$ Fault on Bus~3.}

\begin{figure}[!t]
\centerline{\includegraphics[width=0.9\columnwidth]{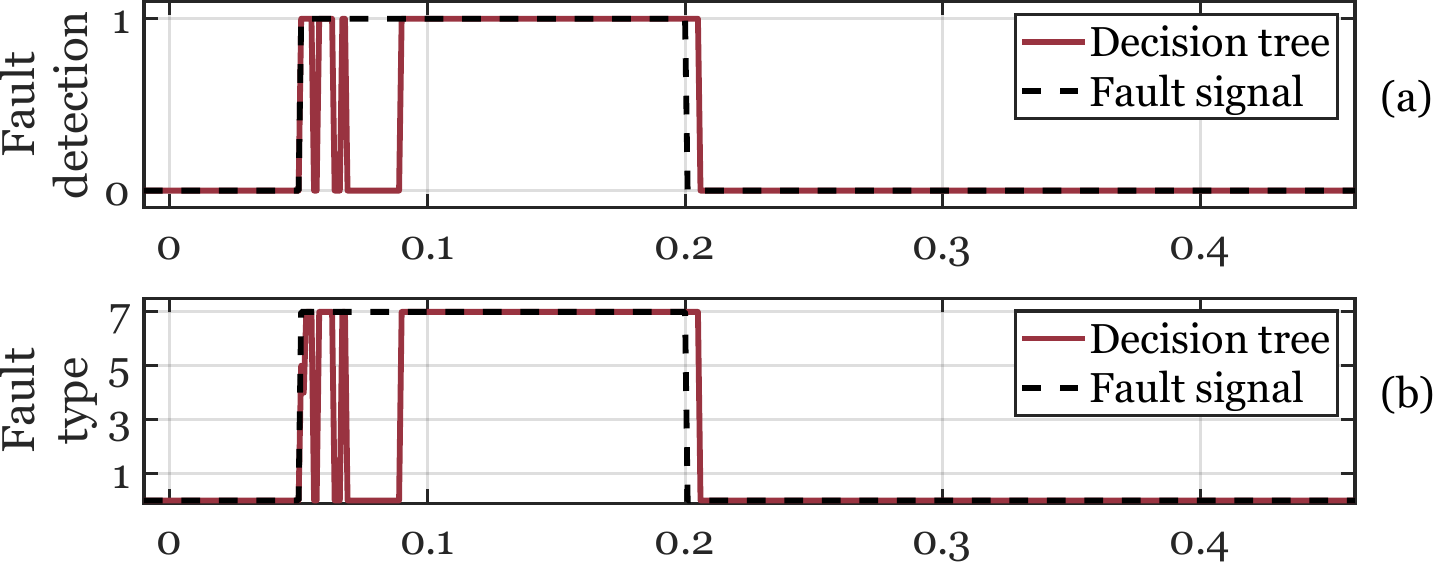}}
 %\vspace*{-0.18 cm}
\caption{ $ABCG$ fault located at bus~3: (a) fault detection and (b) fault type.}
\label{case3}
\end{figure}

%Figs.~\ref{case3}(a)--(c) show fault detection, fault type, and relay RMS current for a three-phase to ground fault with a 40~$\Omega$ impedance located at bus~3, starting at $t=50$~ms and clearing at $t=200$~ms.
Figs.~\ref{case3} shows fault detection and fault type for a three-phase to ground fault with a 40~$\Omega$ impedance located at bus~3, starting at $t=50$~ms and clearing at $t=200$~ms.
Unlike Case~1 and Case~2, the decision tree has not been trained for a 40~$\Omega$ fault value, decreasing its accuracy.
The decision tree detects the fault occurrence at $t=50$~ms and fault clearance at $t=210$~ms; however, it cannot detect correctly after the fault occurrence in a few time steps shown in Fig.~\ref{case3}(a). Furthermore, the decision tree has difficulty identifying the fault type at the first 40~ms of the simulation; however, it can detect the fault with certainty after that. 
The decision tree detects the fault clearance with a 10~ms delay.
The decision tree indicates that the fault type number is 7 after 40~ms, which is equivalent to an $ABCG$ fault using shown in Fig.~\ref{case3}(b).
%Fig.~\ref{case4}(c) shows the current, and all three-phase currents increase since the fault is an $ABCG$ fault.
Fig.~\ref{case3} indicates that the proposed method still works well for an untrained high-impedance fault.
%

%\subsubsection{ Case4: $BG$ Fault on Bus~4.}

%\begin{figure}[!t]
%\centerline{\includegraphics[width=0.95\columnwidth]{figs/Res_100_FTyp_2_Dur_0.05_BusNum_4_KPEC.pdf}}
%\caption{ $BG$ fault at bus~4 with a 100~$\Omega$ resistance: (a) fault detection, (b) fault type, (c) relay RMS current.}
%\label{case4}
%\end{figure}

%Figs.~\ref{case4}(a)--(c) show fault detection, fault type, and relay RMS current for phase $B$ to ground fault with a 100~$\Omega$ resistance located at bus~4, starting at $t=50$~ms and clearing at $t=100$~ms.
%Fig.~\ref{case4}(a) shows the decision tree detecting the fault at $t=50$~ms and detecting fault clearance at $t=101$~ms. The decision tree detects the fault occurrence and clearance accurately.
%Fig.~\ref{case4}(b) shows the decision tree identifying the fault type number correctly during the whole fault period, except at $t=51$~ms. However, it corrects its prediction in the next time step.
%Fig.~\ref{case4}(c) shows the decision tree fault current where phase $B$ current shows the highest increase since the fault is $BG$.
%This case indicates the efficacy of the proposed algorithm in high-impedance faults. 
%%%

\section{Conclusion}

Low fault current and bidirectional current are major challenges in the protection of fully inverter-based microgrids. This work proposes Decision tree--based protection solutions that incorporate fault detection and identification. These solutions are designed for efficient implementation in digital relays due to their simplicity.
The effectiveness of the proposed method is examined for both low and high-impedance faults across seven distinct fault types characterized by varying fault resistance. The analyses are conducted in the context of a 100\% inverter-based microgrid equipped with four inverters.
%The fault is detected in less than 5~ms.
This work indeed facilitates the utilization of a ML-based method for the protection of fully inverter-based microgrids. Future work includes investigating the efficacy of the ML-based methods for a larger microgrid and assessing their accuracy under various network configurations.

\bibliography{IEEEabrv,ref}
\bibliographystyle{IEEEtran}

\vspace{12pt}

\end{document}